\documentclass[a4paper,10pt]{appolb}
\usepackage{graphicx}
\usepackage{float}
\usepackage{epstopdf}
\usepackage{amsmath,amssymb}
\usepackage{verbatim}

\title{
First Results of a Detailed Analysis of \\ p+p Elastic Scattering Data 
from ISR to LHC Energies \\ in the Quark-Diquark Model
\footnote{Presented by F. Nemes at the Wilhelm and Else Heraeus SummerSchool on Diffractive and Electromagnetic Processes at High Energies, Heidelberg, Germany, September 5-9, 2011}}

\begin{document}
\pagestyle{plain}

\author{F. Nemes 
\address{Institute of Physics, E\"otv\"os University,  \\
P\'azm\'any P. s. 1/A, H-1117 Budapest, Hungary\\ 
    frigyes.janos.nemes@cern.ch}
\and
T. Cs\"org\H{o}
\address{Wigner RCP, RMKI, H-1525 Budapest 114, P.O.Box 49, Hungary \\ csorgo.tamas@wigner.mta.hu}
}

\maketitle

\begin{abstract}
First results of a detailed analysis of p+p elastic scattering data
are presented from ISR to LHC energies utilizing the quark-diquark model
of protons in a form  proposed by Bialas and Bzdak.
The differential cross-section of elastic proton-proton collisions 
is analyzed in detailed and systematic manner at small momentum transfers,
starting from the energy range of CERN ISR at  $\sqrt{s}= 23.5$ GeV,
including also recent TOTEM data at the present LHC energies at $\sqrt{s} = 7$ TeV. 
These studies confirm the picture that the size of protons increases systematically 
with increasing energies, while the size of the constituent quarks and diquarks
remains approximately independent of (or only increases only slightly with) the colliding energy. 
The detailed analysis indicates correlations between model parameters and also indicates 
an increasing role of shadowing at LHC energies.
\end{abstract}

\newpage

\section{Introduction}
The differential cross section of elastic scattering of p+p collisions allows one to 
study the internal structure of protons using the theory of diffraction. 
Varying the momentum transfer one can change the resolution of the investigation: 
increasing the momentum transfer corresponds to looking more and more deeply inside the
structure of protons. One of the fundamental outcomes of diffractive p+p scattering
studies was the indication that protons have a finite size and a complicated
internal structure, thus the protons can be considered as composite objects.

Our interest in this problem has been triggered by two factors:
an interesting series of recent theoretical work and also new data from
the TOTEM experiment at CERN LHC. These are detailed below. 
Recently, we became aware of an 
inspiring series of papers of Bialas, Bzdak and collaborators, who studied
elastic proton-proton~\cite{Bialas:2006qf}, 
pion-proton~\cite{Bzdak:2007qq}
and nucleus-nucleus collisions~\cite{Bialas:2006kw},\cite{Bialas:2007eg}
in a framework where the proton was considered as a composite object that contains
correlated quark and diquark constituents. 
In this work, we confirm their main conclusion:
the quark-diquark model of nucleon structure at low momentum transfer does 
capture the main features of this problem and indeed it deserves a closer, more detailed attention.
This study is dedicated to a follow-up, more detailed investigation,
not only including an estimation of the best values of  the parameters,
as was done in ref.~\cite{Bialas:2006qf}, 
but also determining their errors and also the evaluation of the models
based on an analysis of the presented fit quality.
In order to reach these goals,
we utilized standard experimental techniques, such as multi-parameter 
optimalization or fitting with the help of the  MINUIT function minimalization
and multi-parameter optimalization package~\cite{James:1975dr}.

In addition to these simple, straightforward and
interesting theoretical investigations of elastic scattering data from 
CERN ISR in the energy range of 
$\sqrt{s} =$ 23.5, 30.7, 52.9 and   62.5 GeV,
that were already analyzed in ref.~\cite{Bialas:2006qf}, 
new elastic scattering data became available recently 
at $\sqrt{s} = 7$ TeV~\cite{Antchev:2011zz}
from the CERN LHC experiment TOTEM.
So we have tested the model of Bialas and Bzdak 
not only at ISR energies but also at the currently available
highest LHC energies on recent TOTEM data,
in order to learn more details about the evolution of the properties 
of p+p elastic interactions in the recently opened, few TeV energy range.

		Most of the arguments for the composite structure of hadrons have been 
derived from the studies of lepton-hadron interactions. The emerging standard
picture is that hadrons are either mesons, composed of valence quarks and anti-quarks,
or (anti)baryons, composed of three valence (anti)quarks, that
carry the quantum numbers, while the electrically neutral gluons
carry color charges and provide the binding among the quarks and anti-quarks. 
The exact contribution of quarks, gluons, and the sea of virtual quark-antiquark pairs and gluons
to certain hadronic properties e.g. spin is still under detailed investigation.
For example, the gluon contribution to the proton spin is still 
not fully constrained~\cite{Adare:2008px}.
Also, more than 10 exotic hadronic resonances called X, Y and Z states were recently discovered
in electron-positron collisions at the world's highest luminosities in the 
BELLE experiment at KEK. These hadronic states 
cannot be interpreted in the standard picture of quark-antiquark 
or three (anti)quark bound states, according to refs.~\cite{KEK-XYZ,Kreps:2009ne}
Thus even nowadays there are still several open questions that are related to the
compositeness of the hadrons in general.
As gluons do not interact with directly with leptons, their properties are best explored
with the help of the strong hadronic interactions. For example, the gluon contribution
to the proton spin is investigated with the help of  polarized proton - polarized proton collisions
at RHIC~\cite{Adare:2008px}.

In this work, we focus on the effects of internal correlations between the quarks
inside the protons, examining in detail proton-proton elastic scattering 
at several ISR and also at the currently maximal available LHC energies.
Let us recall, that a similar analysis involving three independent quarks 
was not able to properly describe the ISR data ~\cite{Bialas:1977xp}.
In that model quarks were considered as ``dressed'' valence quarks 
in the sense that they contain the gluonic and $q\bar{q}$ contribution as well, as if 
the glue would be concentrated around pointlike valence quarks. 
		30 years after the three independent quark model of ref.~\cite{Bialas:1977xp}
        another three-quark model of the protons was proposed in ref.~\cite{Bialas:2006qf}, 
        that included interesting correlations between 
        two dressed valence quarks to form a diquark. This quark-diquark model~\cite{Bialas:2006qf}, 
        is the basis of our current study,  in order to examine in details
        proton-proton elastic scattering at ISR and LHC energies. 

The quark-diquark  picture of elastic p+p scattering 
resembles to the Glauber optical model~\cite{Glauber:1955qq} 
in nuclear physics, where a multiple expansion is applied. 
The Glauber model, developed originally for nuclear multiple scattering 
problems like cross sections of protons and neutrons on deuteron, 
became a standard model of high energy interactions in nuclear physics
where multiple interactions are built up from superpositions
of nucleon-nucleon scattering. This model became a fundamental and
successfull tool in describing nuclear collisions at high energy ~\cite{Glauber:2006gd}. 

\par

   The body of this manuscript is organized as follows: 
   the theoretical models are described in Section \ref{sec:theory}, 
   for both cases, when the diquark acts as a single entity and also when it acts  
   as a composite object. 
   In Section \ref{sec:reproduction} the original results of Bialas and Bzdak 
   are reproduced and detailed, including a note on the quality of these data descriptions.
   Section \ref{sec:minuit}
   contains our main new results, the first  MINUIT results. The conclusion takes place in Section \ref{sec:conclusion}.

\newpage

\section{Elastic scattering in the quark-diquark model}
\label{sec:theory}

\begin{figure}[H]
    	\includegraphics[width=0.6\textwidth]{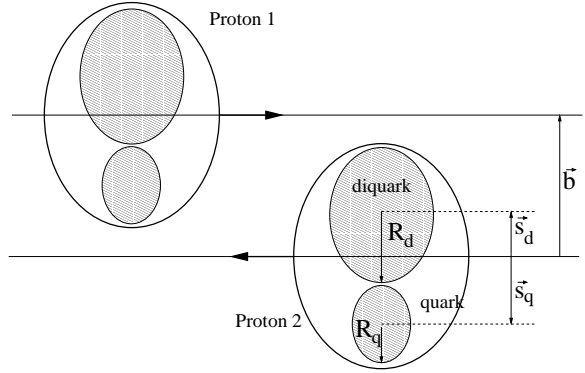}
		\centering
		\caption{The schematic scattering situation of the two protons when the diquark is assumed to be a single entity.}
		\label{scattering sit1}
    \end{figure}

	We describe proton-proton interactions as collision of two systems, each one composed of a dressed quark and diquark. The collision is schematically illustrated
	on Fig. \ref{scattering sit1}.

	The interaction between quarks and diquarks is assumed to be purely absorptive. Consequently the amplitude has no real part and the imaginary part -- 
	dominating at high energy -- is given by the absorption of the incoming particle wave, namely the inelastic (non-diffractive) collisions.\par
    In the impact parameter space the inelastic proton-proton cross-section for a fixed impact parameter $\vec{b}$ can be given by the following formula \cite{Bialas:2006qf}
	\begin{equation}
		\sigma(\vec{b})=\int\limits^{+\infty}_{-\infty}...\int\limits^{+\infty}_{-\infty}{\text{d}^2s_q \text{d}^2s'_q \text{d}^2s_d \text{d}^2s'_d D(\vec{s_q},\vec{s_d})
		D(\vec{s_q}',\vec{s_d}')
		\sigma(\vec{s_q},\vec{s_d};\vec{s_q}',\vec{s_d}';\vec{b})},
		\label{elsoegyenlet}
	\end{equation}
	where $\vec{s_{q}}$, $\vec{s_{q}}'$ and $\vec{s_{d}}$, $\vec{s_{d}}'$ are the transverse positions of the quarks and diquarks respectively, and
	 the integrand is the probability function 
	of having inelastic interaction at a given impact parameter $\vec{b}$ and transverse	positions of the constituents. \par 
	The quark-diquark distribution inside the nucleon is taken into account with the following Gaussian
    \begin{equation}
    	D\left(\vec{s_q},\vec{s_d}\right)=\frac{1+\lambda^2}{\pi R_p^2}e^{-(s_q^2+s_d^2)/R_p^2}\delta^2(\vec{s_d}+\lambda \vec{s_q}),\;\lambda=m_q / m_d,
    \end{equation}
	where  $R_p$ is the ``proton size'', the variance of the distribution and $\lambda$ is the mass ratio of the quark and the diquark. Obviously $ 1/2 \le \lambda \le 1$, 
	where $1/2$ would indicate a loosely bound diquark. The two dimensional delta function preserves the center of mass in the tranverse plane.\par
	Elastic interactions of the constituents are {\it independent} inside the proton, accordingly the probability distribution of elastic proton-proton collision is
    the {\it product} of the probability distribution of elastic interactions of the constituents \cite{Glauber,Czyz:1969jg}
	\begin{align}
		\sigma(\vec{s_q},\vec{s_d};\vec{s_q}',\vec{s_d}';\vec{b})=1-\prod_{a,b \in \{q,d\}}\{1-\sigma_{ab}(\vec{b} + \vec{s_a}' - \vec{s_b}' )\}.
	\label{Glauber expansion}
	\end{align}
	The inelastic differential cross-sections are parametrized with Gaussian distributions
	\begin{equation}
		\sigma_{ab}\left(\vec{s}\right) = A_{ab}e^{-s^2/R_{ab}^2},\;R_{ab}^2=R_a^2+R_b^2,
		\label{inelastic cross sections}
	\end{equation}
	where $R_{ab}$ is the variance of having an inelastic collision, which is calculated from the sum of the squared $R_q$, $R_d$ radius parameters; the $A_{ab}$ parameters
	are the ampitudes. From unitarity the elastic 
	amplitude in impact parameter space\footnote{As it was mentioned the real part of the amplitude is ignored.}
		\begin{equation}
			t_{el}(\vec{b})=1-\sqrt{1-\sigma(\vec{b})}.
		\end{equation}
	The elastic amplitude in momentum transfer representation is the Fourier-transform of the amplitude in impact parameter space 
		\begin{equation}
			T(\vec{\Delta})=\int\limits^{+\infty}_{-\infty}\int\limits^{+\infty}_{-\infty}{t_{el}(\vec{b})e^{i\vec{\Delta} \cdot \vec{b}}\text{d}^2b}=
			2\pi\int\limits_0^{+\infty}{t_{el}\left(b\right)J_0\left(\Delta b\right)b {\text d}b},
		\end{equation}
	where $\Delta=|\vec{\Delta}|$, $b=|\vec{b}|$ and $J_0$ is the zeroth Bessel-function of the first kind. Then the elastic differential cross section reads as
	\begin{equation}
		\frac{d\sigma}{dt}=\frac{1}{4\pi}\left|T\left(\Delta\right)\right|^2.
	\end{equation}

%\newpage

\subsection{The diquark is assumed to act as a single entity}
	
	The subject of this section is to analyse that case when the quark and diquark radii are independent model parameters, which means that the diquark is considered as one entity as
	indicated on Fig. \ref{scattering sit1}. In
    this case, the number of free parameters can be reduced if we assume that the number of partons is twice as many in the diquark than in the quark. From the inelastic differential 
    cross sections (\ref{inelastic cross sections}) the total inelastic cross sections are
	\begin{equation}
	\label{totalinelastic}
		\sigma_{ab}=\int\limits^{+\infty}_{-\infty}\int\limits^{+\infty}_{-\infty}{\sigma_{ab}\left(\vec{s}\right)}\text{d}^2s= \pi A_{ab}R_{ab}^2,\; a,b \in \{q,d\}.
	\end{equation}
	Our assumption tells us that 
	\begin{equation}
		\sigma_{qq}:\sigma_{qd}:\sigma_{dd}=1:2:4, 
		\label{ratiosforsigma}
	\end{equation}
	from which we can deduce the following expressions
		\begin{equation}
			A_{qd}=A_{qq}\frac{4R_q^2}{R_q^2+R_d^2},\;A_{dd}=A_{qq}\frac{4R_q^2}{R_d^2},
		\end{equation}
	which means that every $A_{ab}$ parameter can be expressed in term of $A_{qq}$. With these ingredients the calculation of (\ref{elsoegyenlet}) reduces
	to Gaussian integrations. The general or master formula for these Gaussian integrals
    is given in the Appendix.

\par
	In the next two subsections the elastic proton-proton data analysis are 
presented in the case when the diquark is assumed to have no internal, more detailed
structure in elastic collisions. 
First, we demonstrate with plots that we reproduce the results  of the
original paper at ISR energies \cite{Bialas:2006qf}. This forms a solid basis 
for imrovement as presented in the subsequent parts of our current study. 
Our MINUIT fits are then presented, utilizing  same ISR data and then the new results 
	are presented at $7$ TeV utilizing new data of the TOTEM experiment~\cite{Antchev:2011zz}.

%\newpage

\subsection{Analysis of the case when the diquark acts as composite object}

\begin{figure}[H]
        \includegraphics[width=0.6\textwidth]{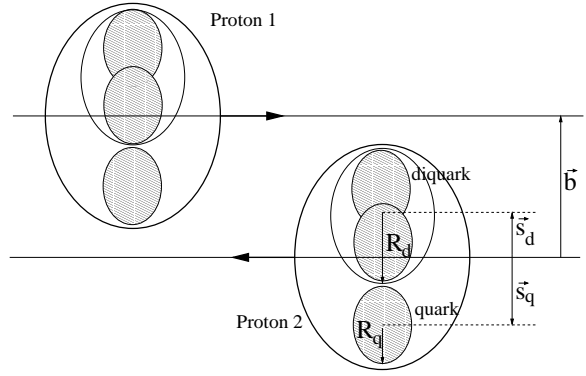}
        \centering
        \caption{The scattering situation of the two protons when the diquark is assumed to be composed of two quarks, and the proton symbolically can be written as p=(q,(qq)).}
        \label{scatter qq}
    \end{figure}

        The scattering when the diquark is a composite object is illustrated on
    Fig. \ref{scatter qq}. The quark distribution inside the diquark is supposed
    to have a Gaussian shape
    \begin{equation}
            D\left(\vec{s_{q1}},\vec{s_{q2}}\right)=\frac{1}{\pi d^2}e^{-\left(s_{q1}^2+s_{q2}^2\right)/2d^2}
        \delta^2\left(\vec{s_{q1}}+\vec{s_{q2}}\right),
        \label{quarkdistribution}
    \end{equation}
    where $\vec{s_{q1}}$ and $\vec{s_{q2}}$ are the transverse quark positions inside the diquark, and
    \begin{equation}
        d^2=R_d^2-R_q^2
    \end{equation}
    is the variance of the quark distribution, calculated from the diquark and quark radius parameters.
        If the diquark has the internal structure (\ref{quarkdistribution}) then the $\sigma_{qd}$,
    $\sigma_{dq}$ and $\sigma_{dd}$ inelastic differential cross sections (\ref{inelastic cross sections})
    can be calculated from $\sigma_{qq}$ using an expansion analogous to expression (\ref{Glauber expansion}). The results
    for $\sigma_{qd}$ and $\sigma_{dd}$ are the following \cite{Bialas:2006qf}
    \begin{equation}
            \sigma_{qd}\left(\vec{s}\right)=\frac{4A_{qq}R^2_q}{R^2_d+R^2_q}e^{-s^2\frac{1}{R^2_d+R^2_q}}-\frac{A^2_{qq}R^2_{q}}{R^2_d}e^{-s^2/R^2_q},
    \end{equation}
    and
            \begin{align}
                \sigma_{dd}\left(\vec{s}\right) & = \frac{4A_{qq}R_q^2}{R_d^2}e^{-s^2\frac{1}{2R_d^2}}-\frac{4A_{qq}^2R_q^4}{R_d^4}e^{-s^2/R_d^2}-\frac{2A_{qq}^2R_q^2}{2R_d^2-R_q^2}e^{-s^2/R_q^2}+ \\
                    & + \frac{4A_{qq}^3 R_q^4}{R_d^2\left(2R_d^2-R_q^2\right)}e^{-s^2\frac{2R_d^2+R_q^2}{2R_q^2 R_d^2}}-\frac{A_{qq}^4R_q^4}{\left(2R_d^2-R_q^2\right)^2}e^{-s^2\frac{2}{R_q^2}} \notag.
            \end{align}
    The relevant formula to calculate the inelastic cross section of eq.(\ref{elsoegyenlet}) is again
    obtained from the expressions summarized in the Appendix.

\newpage

\section{Reproduction and cross-checks of the fits of Bialas and Bzdak}
\label{sec:reproduction}

\subsection{The diquark acts as a single entity}

   	In this section we reproduce the fitting formula of ref.~\cite{Bialas:2006qf}, based on
    the best values of the model parameters that they have published. 
    These values were obtained in ref.~\cite{Bialas:2006qf} from fitting four essential
    elements of the elastic scattering cross-sections: they adjusted the model parameters 
        so that (i) the total inelastic cross-section, (ii) the slope of the differential 
        inelastic cross-section at $t= 0$, (iii) the position of the diffractive minimum, 
        and (iv) the height of the first diffractive maximum just after the minimum
        be in agreement with the data. This method lead to remarkable simplicity in the fitting
        strategy and a very nice, apparent overall agreement with the measured data,
        as can be seen from the Figures of ref.~\cite{Bialas:2006qf} .
    As a consequence of this method, 
    the errors of the fit parameters in ref.~\cite{Bialas:2006qf} 
    were not presented, and the overall  fit quality parameters ($\chi^2$/NDF, CL) 
    were not specified as well, although these are the qualifyers that determine
    the acceptability of certain models when the language of mathematical statistics
    is utilized to characterize the data description. By recalculating the original curves,
    we check in this section the quality of our reproduction of the results of Bialas and Bzdak,
    and also supplemented their paper with the previously undetermined fit quality parameters. 
   
\begin{figure}[H]
                \includegraphics[width=0.9\textwidth]{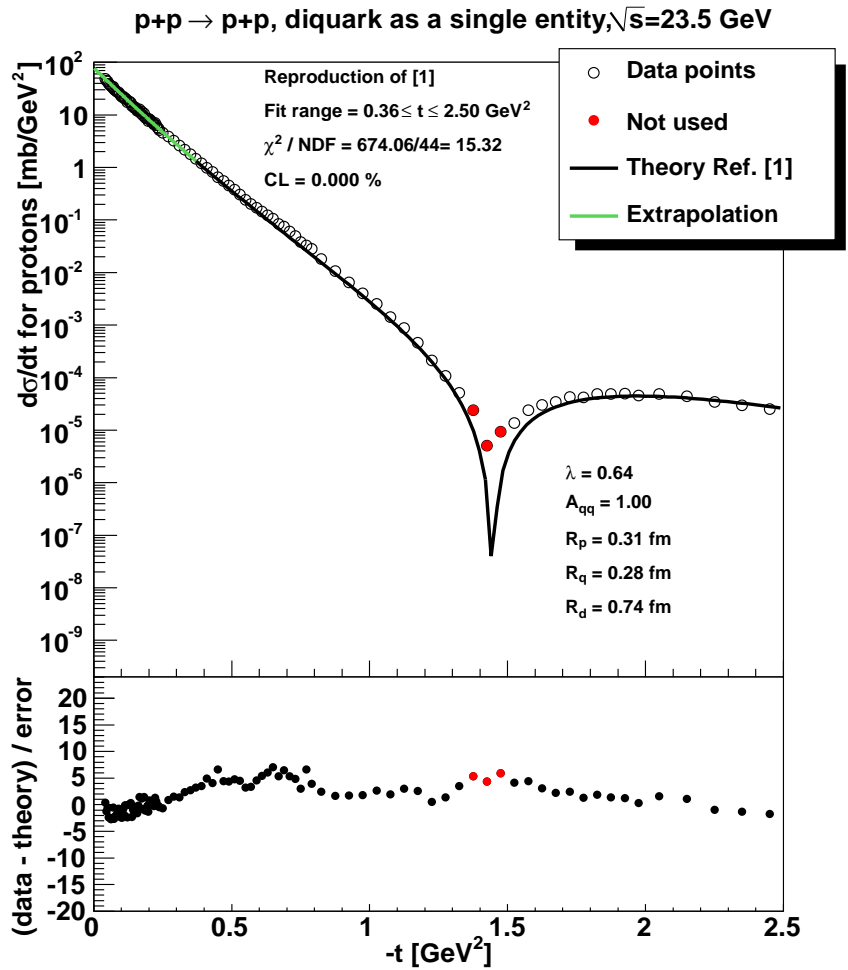}
				\centering
				\caption{Reproduction of the calculation of A. Bialas and A. Bzdak from \cite{Bialas:2006qf} at 23.5 GeV, in the case when the diquark is assumed to be a single entity. 
                We have fixed the values of the model parameters to values given in Ref. \cite{Bialas:2006qf}. In order to determine fit quality, we restricted the fit 
				range to 0.36 GeV $<$ -t $<$ 2.5 GeV, to allow a systematic comparision with more recent TOTEM data at LHC. As the model of Bialas and Bzdak is known
                to be singular at the dip, in this and all other data analysed we have left out 3 data points from the dip region.}
				\label{Bialas1}
            \end{figure}

\begin{figure}[H]
                \includegraphics[width=0.9\textwidth]{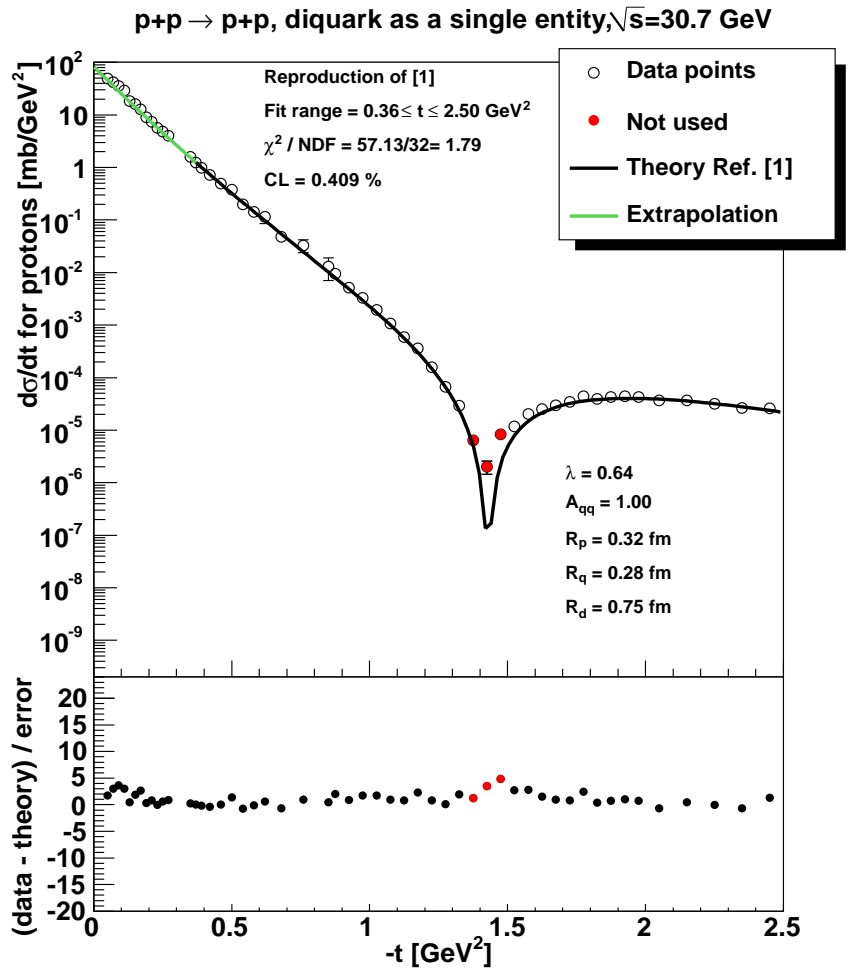}
                \centering
                \caption{Same as Fig. \ref{Bialas1}, but at $\sqrt{s}$=30.7 GeV.}
                \label{Bialas2}
            \end{figure}

\begin{figure}[H]
                \includegraphics[width=0.9\textwidth]{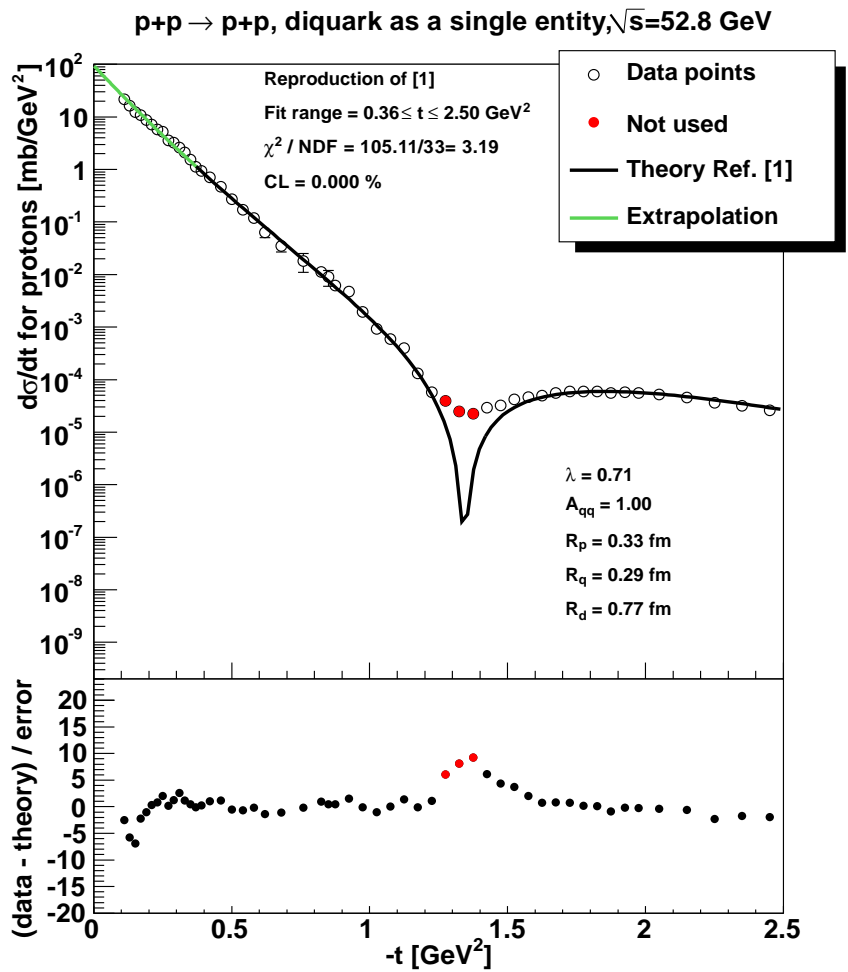}
                \centering
                \caption{Same as Figs. \ref{Bialas1} \& \ref{Bialas2}, at $\sqrt{s}$=52.8 GeV.}
                \label{Bialas3}
            \end{figure}

\begin{figure}[H]
                \includegraphics[width=0.9\textwidth]{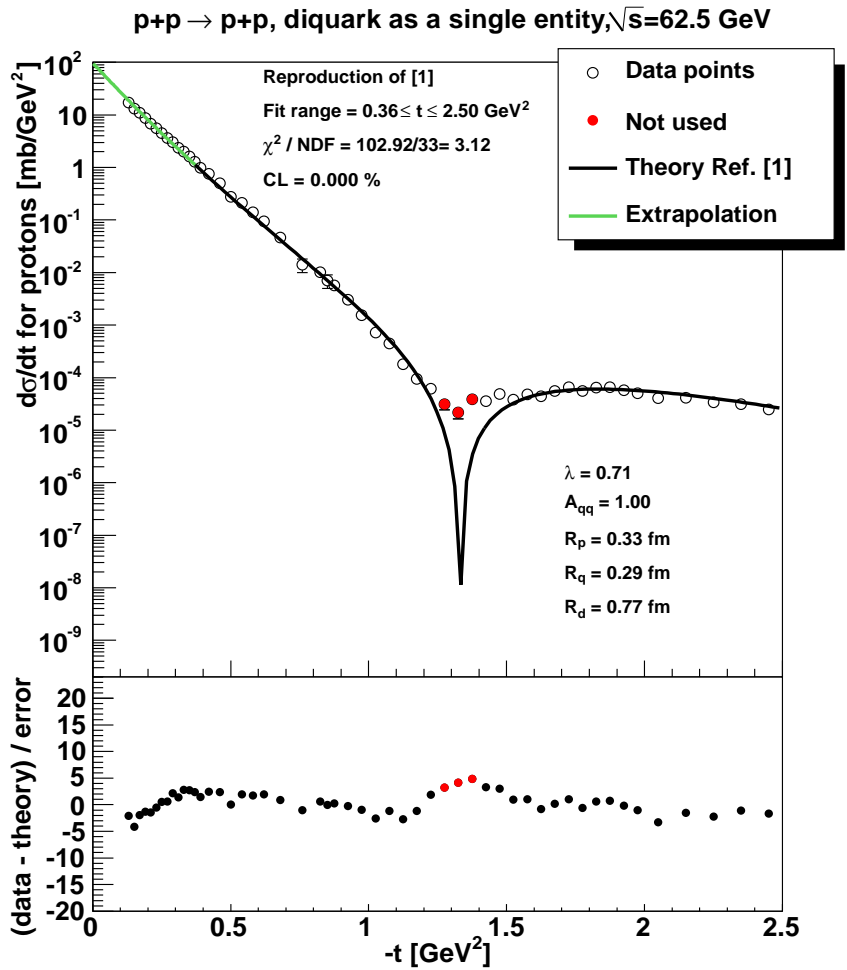}
                \centering
                \caption{Same as Figs. \ref{Bialas1}-\ref{Bialas3}, except that the energy is $\sqrt{s}$=62.5 GeV.}
                \label{Bialas4}
            \end{figure}

The results are shown on Figs. \ref{Bialas1}-\ref{Bialas4} for the energies $\sqrt{s}=$23.5, 30.7, 52.8, 62.5 GeV respectively. More recently the elastic proton-proton $d\sigma/dt$ 
    was measured by the TOTEM experiment at LHC. The first TOTEM
    data were published in the $\left|t\right|$ range from 0.36 GeV up to 2.5 GeV~\cite{Antchev:2011zz}. The fit quality parameters presented here are 
    restricted to those data points which fall into this $\left|t\right|$ region, in order to compare the accuracy of the model in the ISR and LHC energy regimes.

	\vfill
\newpage

\newpage

\subsection{Reproduction of earlier results for the case of composite diquarks }

\begin{figure}[H]
                \includegraphics[width=0.9\textwidth]{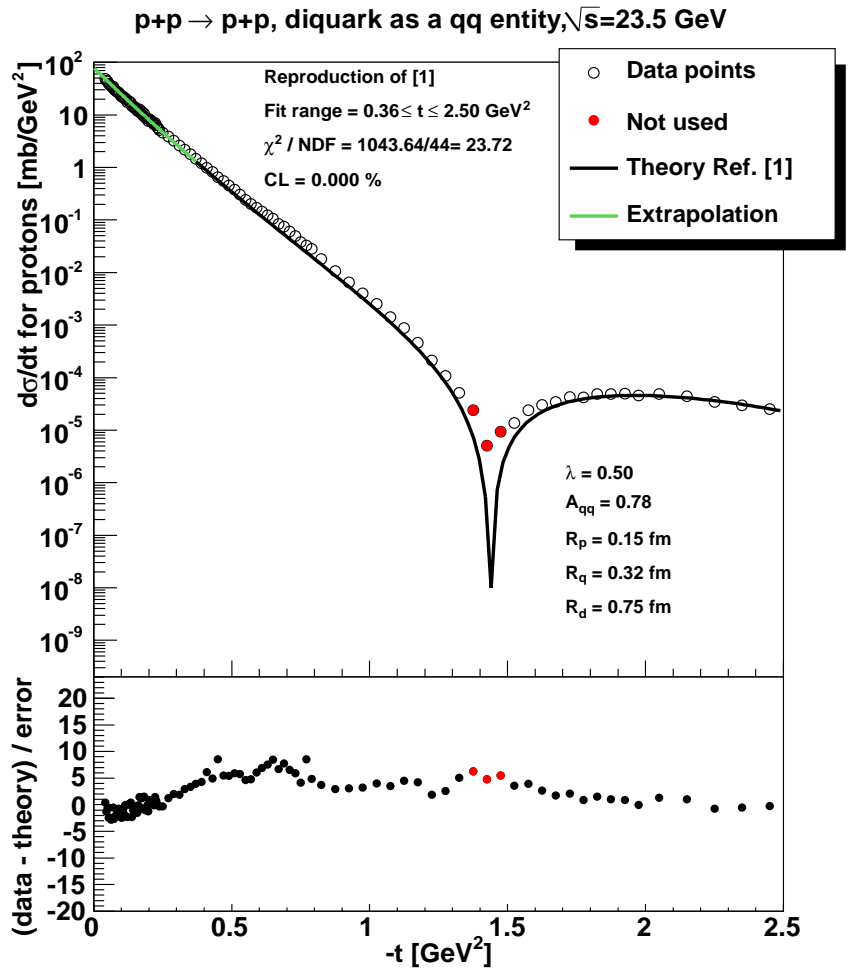}
                \centering
                \caption{
				Reproduction of the calculation of A. Bialas and A. Bzdak \cite{Bialas:2006qf} at 23.5 GeV, in the case when the diquark is assumed to be a qq entity. 
                We have fixed the values of model parameters to values given in \cite{Bialas:2006qf}. In order to determine fit quality, we restricted the fit
                range to 0.36 GeV $<$ -t $<$ 2.5 GeV, to allow a systematic comparision with more recent TOTEM data at LHC. As the model of Bialas and Bzdak is known
                to be singular at the dip, in this and all other data analysed we have left out 3 data points from the dip region.
               }

                \label{Bialas5}
            \end{figure}

\begin{figure}[H]
                \includegraphics[width=0.9\textwidth]{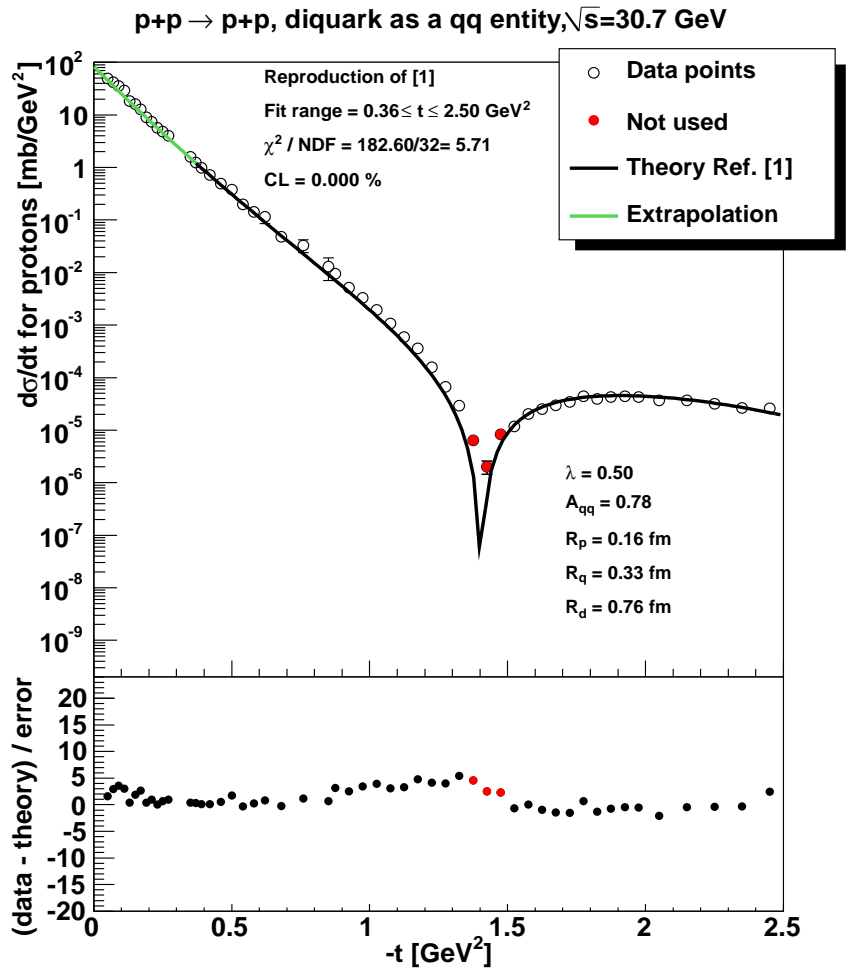}
                \centering
                \caption{ Same as Fig \ref{Bialas5}, except that the analyzed energy is $\sqrt{s}=$ 30.7 GeV.}
                
                \label{Bialas6}
            \end{figure}

\begin{figure}[H]
                \includegraphics[width=0.9\textwidth]{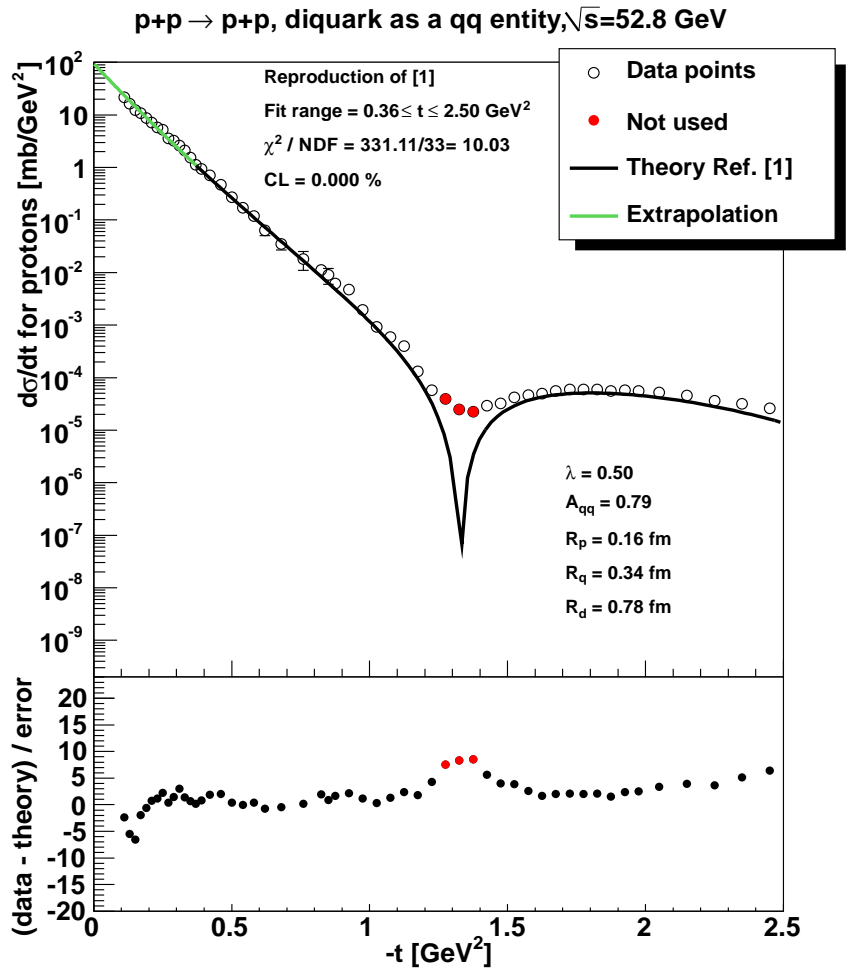}
                \centering
                \caption{Same as Fig \ref{Bialas5}, given at $\sqrt{s}=$ 52.8 GeV.
                }
                \label{Bialas7}
            \end{figure}

\begin{figure}[H]
                \includegraphics[width=0.9\textwidth]{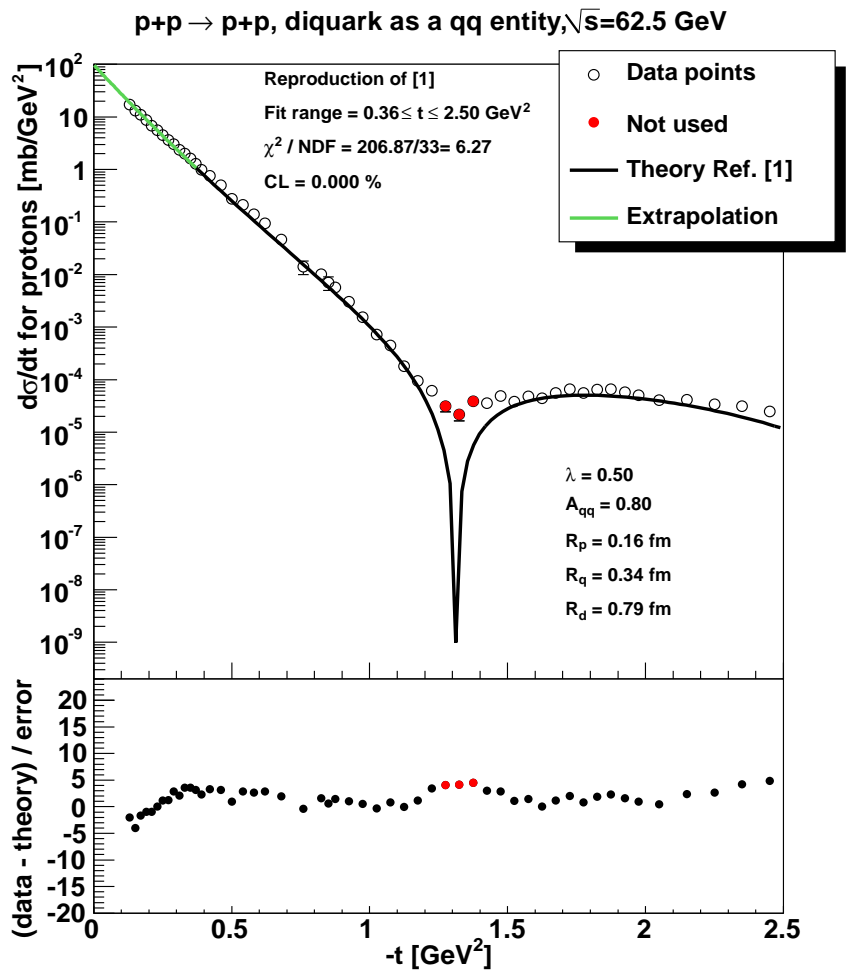}
                \centering
                \caption{Identical with Fig \ref{Bialas5}, except that it is evaluated at $\sqrt{s}=$ 62.5 GeV.
                }
                \label{Bialas8}
            \end{figure}

    We checked the correctness of our implementation of the formula from the original paper \cite{Bialas:2006qf}. In the original
    analysis the errors of the fit parameters were not shown and the overall fit quality parameters ($\chi^2$/NDF, CL) are
    missing as well. The missing fit quality parameters will be supplemented in this section. The results are shown on Fig. \ref{Bialas5}-\ref{Bialas8} for
    the energies 23.5, 30.7, 52.8, 62.5 GeV respectively.
	More recently the elastic proton-proton $d\sigma/dt$
    was measured by the TOTEM experiment at LHC. The first TOTEM
    data covers the $\left|t\right|$ range from 0.36 GeV up to 2.5 GeV~\cite{Antchev:2011zz}. The fit quality parameters presented here are
    restricted to those data points which fall into this region, in order to compare the accuracy of the model in the ISR and LHC energy regime.
\vfill
\newpage

\newpage

\section{MINUIT fit results for ISR and TOTEM data}
\label{sec:minuit}

	\subsection{The diquark acts as a single entity}

\begin{figure}[H]
		\includegraphics[width=0.9\textwidth]{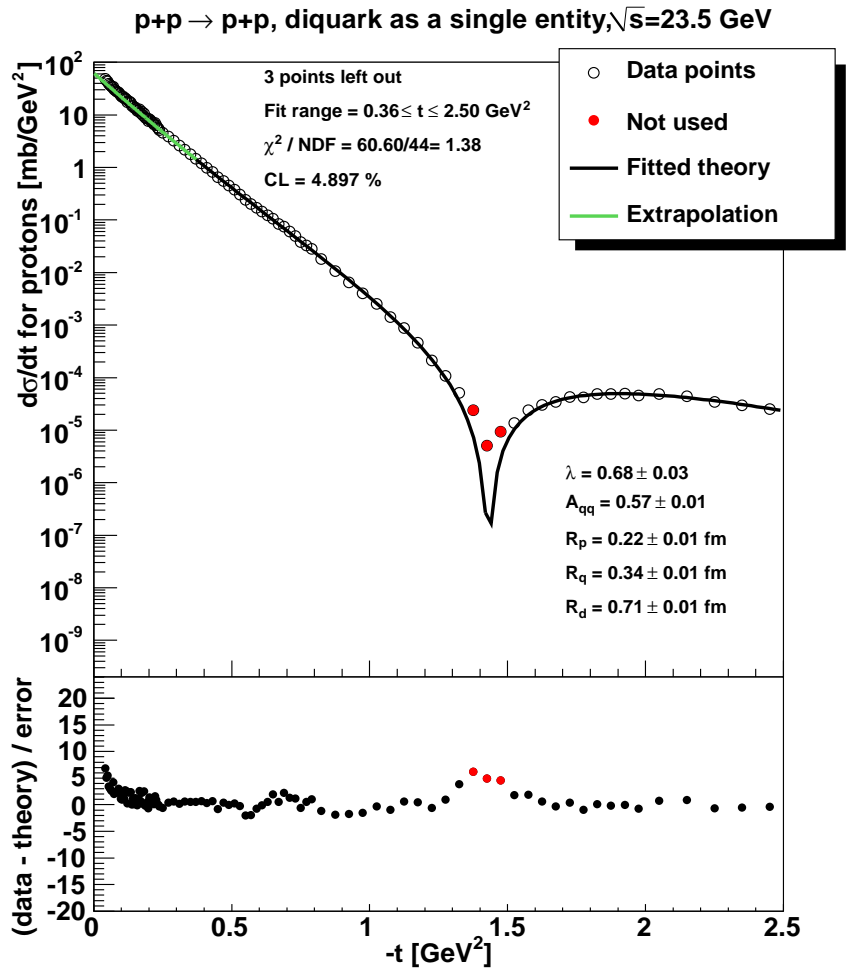}
        	\centering
                \caption{Results of MINUIT fits at ISR energies when the diquark is assumed to be a single entity.  The confidence level is higher than 0.1\%, which means that the 
				fit quality is acceptable.
				The fit was mode on the 0.36 - 2.5 GeV $\left|t\right|$ range according to ~\cite{Antchev:2011zz}.
				As the model is singular around the dip, 3 data points at the dip were left out from the fit.
				The parameter values are given with statistical errors. The systematic errors are not yet investigated and the effects of the correlation between 
				model parameters are not yet determined, in this sense this result is still preliminary.}
		\label{singlefitfor23}
	\end{figure}

\begin{figure}[H]
					\includegraphics[width=0.9\textwidth]{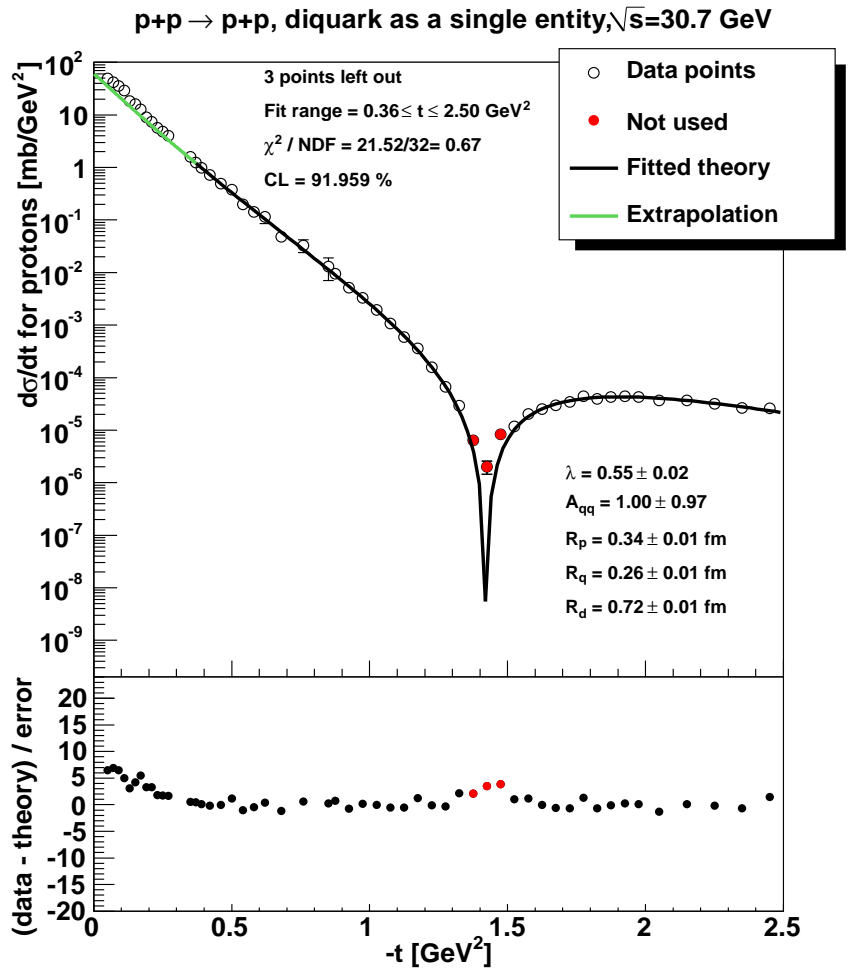}
                    \centering
                    \caption{Same as Fig. \ref{singlefitfor23}, but for the energy $\sqrt{s}=30.7$ GeV. }
			\label{singlefitfor31}
                \end{figure}

\begin{figure}[H]
					\includegraphics[width=0.9\textwidth]{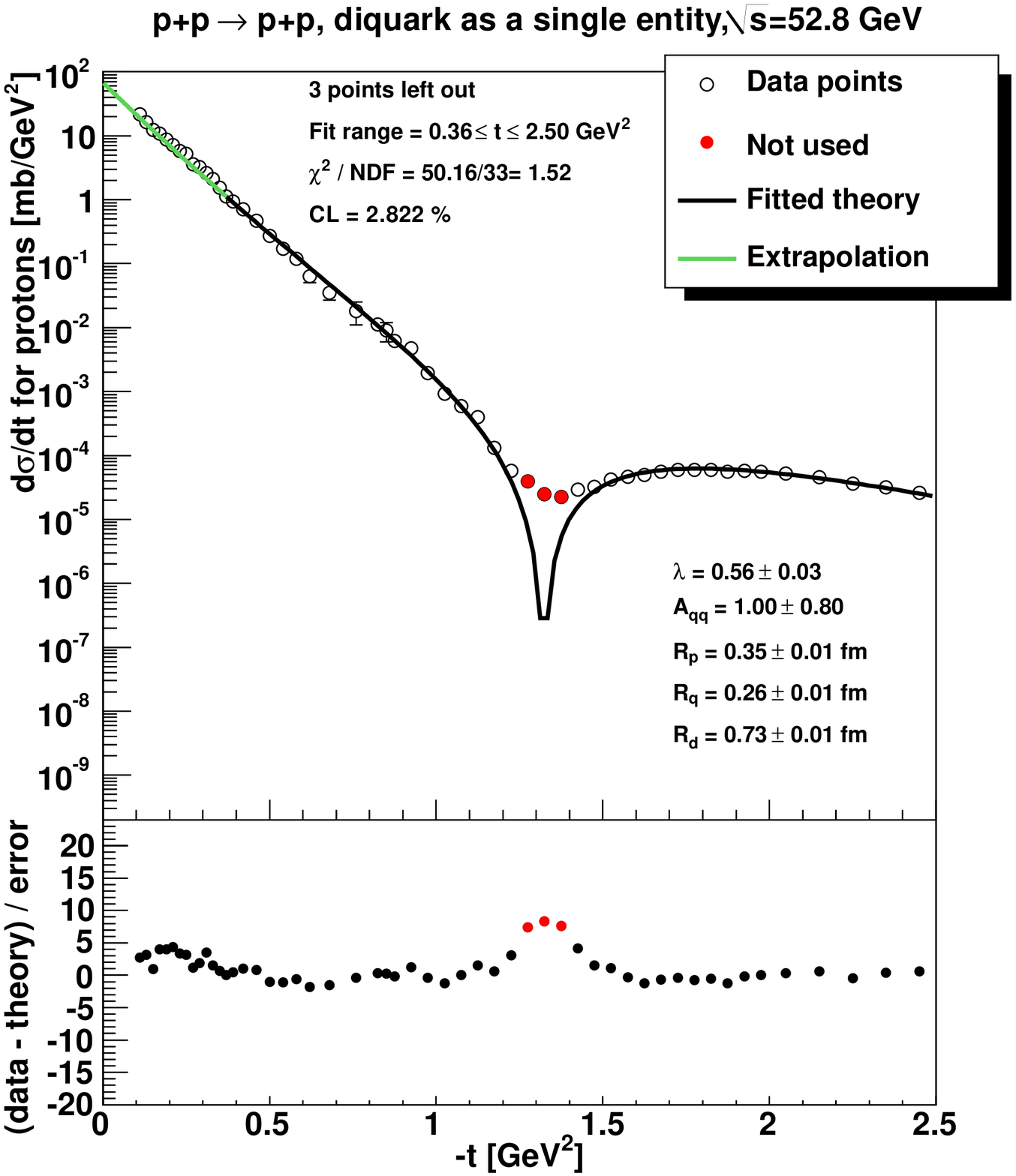}
                    \centering
                    \caption{Same as Fig. \ref{singlefitfor23}, but for the energy $\sqrt{s}=52.8$ GeV. }
			\label{singlefitfor53}
	\end{figure}

\begin{figure}[H]
			\includegraphics[width=0.9\textwidth]{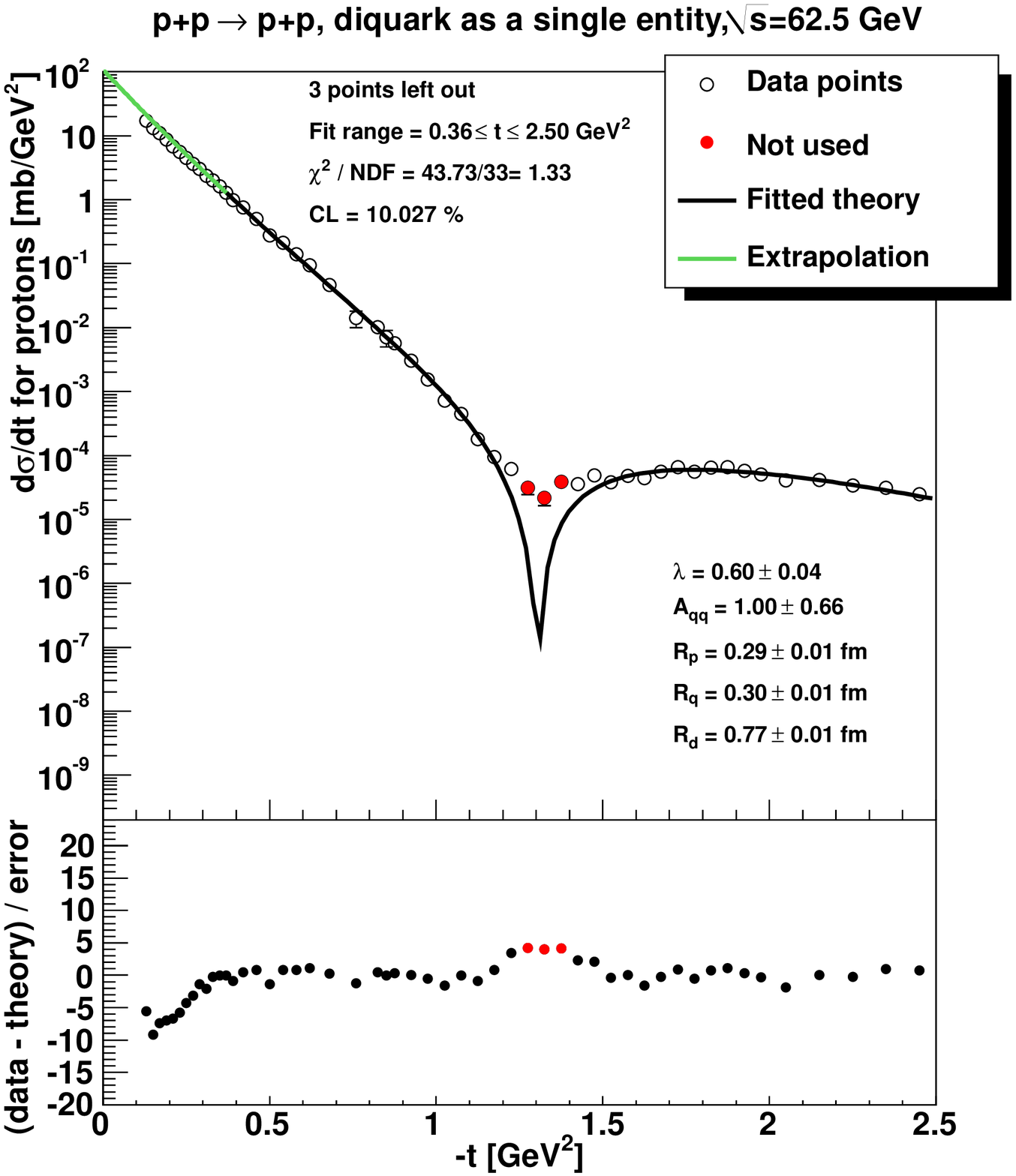}
			\centering
			\caption{Same as Fig. \ref{singlefitfor23}, but for the energy $\sqrt{s}=62.5$ GeV.}
		\label{singlefitfor62}
	\end{figure}

\begin{figure}[H]
                \includegraphics[width=0.9\textwidth]{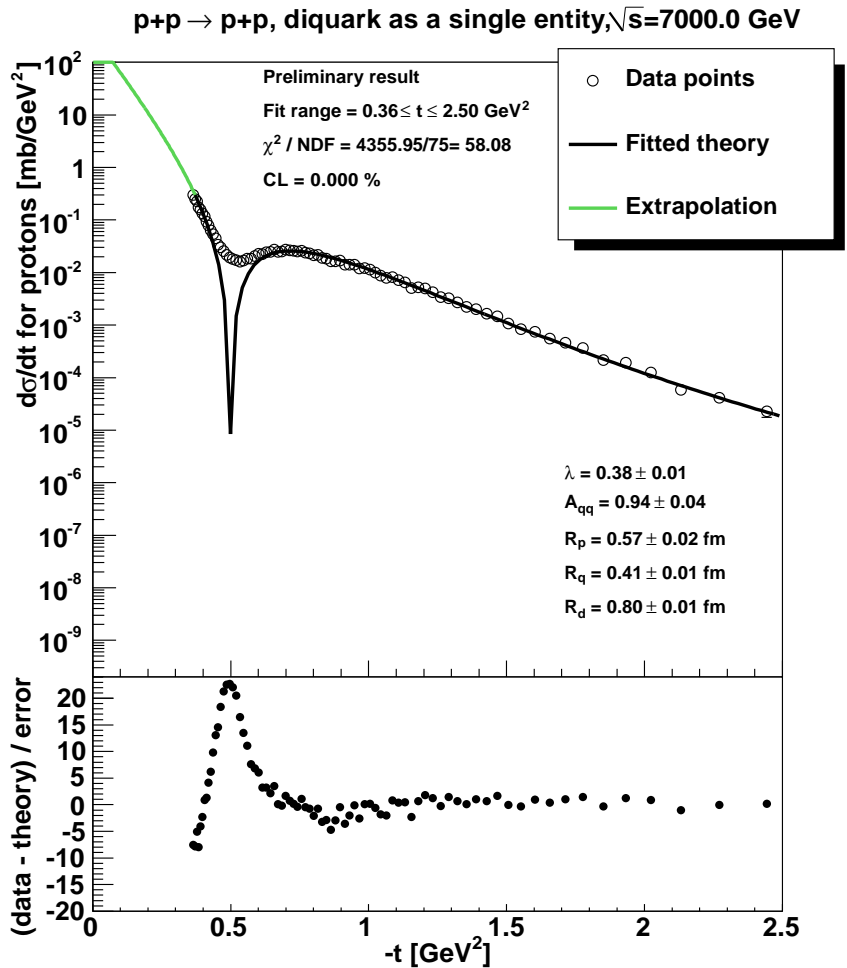}
                \centering
		\caption{The result of the fit at LHC at $7$ TeV when the diquark is assumed to be one entity. The obtained parameter values are given
				with statistical errors. Note that CL is below 0.1\%, so the quality of this fit is
                not acceptable.The systematic errors are not yet investigated and the effects of the correlation between
                model parameters are not yet determined, in these sense this result is still preliminary.}
		\label{singlefittotems}

	\end{figure}

	In this section the MINUIT fit results are presented for the 
    ISR \cite{Nagy:1978iw,Amaldi:1979kd} and TOTEM~\cite{Antchev:2011zz} proton-proton elastic scattering 
	data using the single entity model to describe the diquark. Results are illustrated on Fig. \ref{singlefitfor23}-\ref{singlefittotems}. The confidence levels, and model parameters
	together with their errors are presented in Table \ref{parameters single}, amended with the calculated total elastic cross sections including their uncertainty evaluated from
    the MINUIT fits. The 
    ratios of the inelastic cross sections were fixed with expression (\ref{ratiosforsigma}) in order to decrease the number
    of free parameters, therefore these ratios will be provided only for the case when the diquark is assumed to be a composite object.\par  
	We have to give some preliminary remarks before
    presenting the result. The Bialas-Bzdak model shows a singular behaviour at the dip position, having no real
    part in the amplitude. In order to give a meaning to the fit in this region 3 data points were left out 
    from this
	dip region during the fit; these points are shown in red on the plots. \par 
	Another important remark is that the TOTEM data covers the $|t|$ range from 0.36 GeV up to 2.5 GeV 
    and this range is applied
    in our minimization procedure to allow a comparison between the ISR and TOTEM results. 
	Note that Bialas and Bzdak adjusted the model to account only for the slope and the value at 
    t = 0 together with the position of the minimum \cite{Bialas:2006qf}. A different
    strategy is followed here, since the theoretical curve was fitted directly to the experimental data points 
    using the CERN MINUIT package.  
\vfill
\newpage

	\begin{table}
	    \begin{tabular}{|c|c|c|c|c|c|} \hline
			$ \sqrt{s}$ [GeV]& 23.5  & 30.7 & 52.9 & 62.5 & 7000   \\ \hline\hline 
			$R_{p}$ [fm] & 0.22 $\pm$ 0.01 &  0.34 $\pm$ 0.01 &  0.35 $\pm$ 0.01 &  0.29 $\pm$ 0.01 &  0.57 $\pm$ 0.02 \\ \hline 
			$R_{q}$ [fm] & 0.34 $\pm$ 0.01 &  0.26 $\pm$ 0.01 &  0.26 $\pm$ 0.01 &  0.30 $\pm$ 0.01 &  0.41 $\pm$ 0.01 \\ \hline 
			$R_{d}$ [fm] & 0.71 $\pm$ 0.01 &  0.72 $\pm$ 0.01 &  0.73 $\pm$ 0.01 &  0.77 $\pm$ 0.01 &  0.80 $\pm$ 0.01 \\ \hline 
			$\lambda$ & 0.68 $\pm$ 0.03 &  0.55 $\pm$ 0.02 &  0.56 $\pm$ 0.03 &  0.60 $\pm$ 0.04 &  0.38 $\pm$ 0.01 \\ \hline 
			$A_{qq}$ & 0.57 $\pm$ 0.01 &  1.00 $\pm$ 0.97 &  1.00 $\pm$ 0.80 &  1.0 $\pm$ 0.66 &  0.94 $\pm$ 0.04 \\ \hline 
			$\chi^2/NDF$& 60.6/44 & 21.5/32 & 50.2/33 & 43.73/33 & 4355.9/75  \\ \hline 
			$CL$ [\%] & 4.9 & 92.0 & 2.8  & 10.0  & 0.0  \\ \hline 
			$\sigma^{elastic}_{total}$ [mb] & 
				6.0  & 
				5.6  & 
				5.9  & 
				8.7  & 
				20.3 \\ \hline 
       	\end{tabular}
		\caption{The overall fit quality and resulting parameters of the fit at ISR energies including 
 	the LHC result at $7$ TeV. The diquark is
             assumed to be a single entity. The obtained parameter values are given with statistical errors. The systematic errors are not yet investigated and the effects of the correlation between
                model parameters are not yet determined, so these results in this sense are preliminary.}
		\label{parameters single}
	\end{table}

%\newpage

\subsection{The diquark is assumed to be a composite object}
\begin{figure}[H]
    	\includegraphics[width=0.9\textwidth]{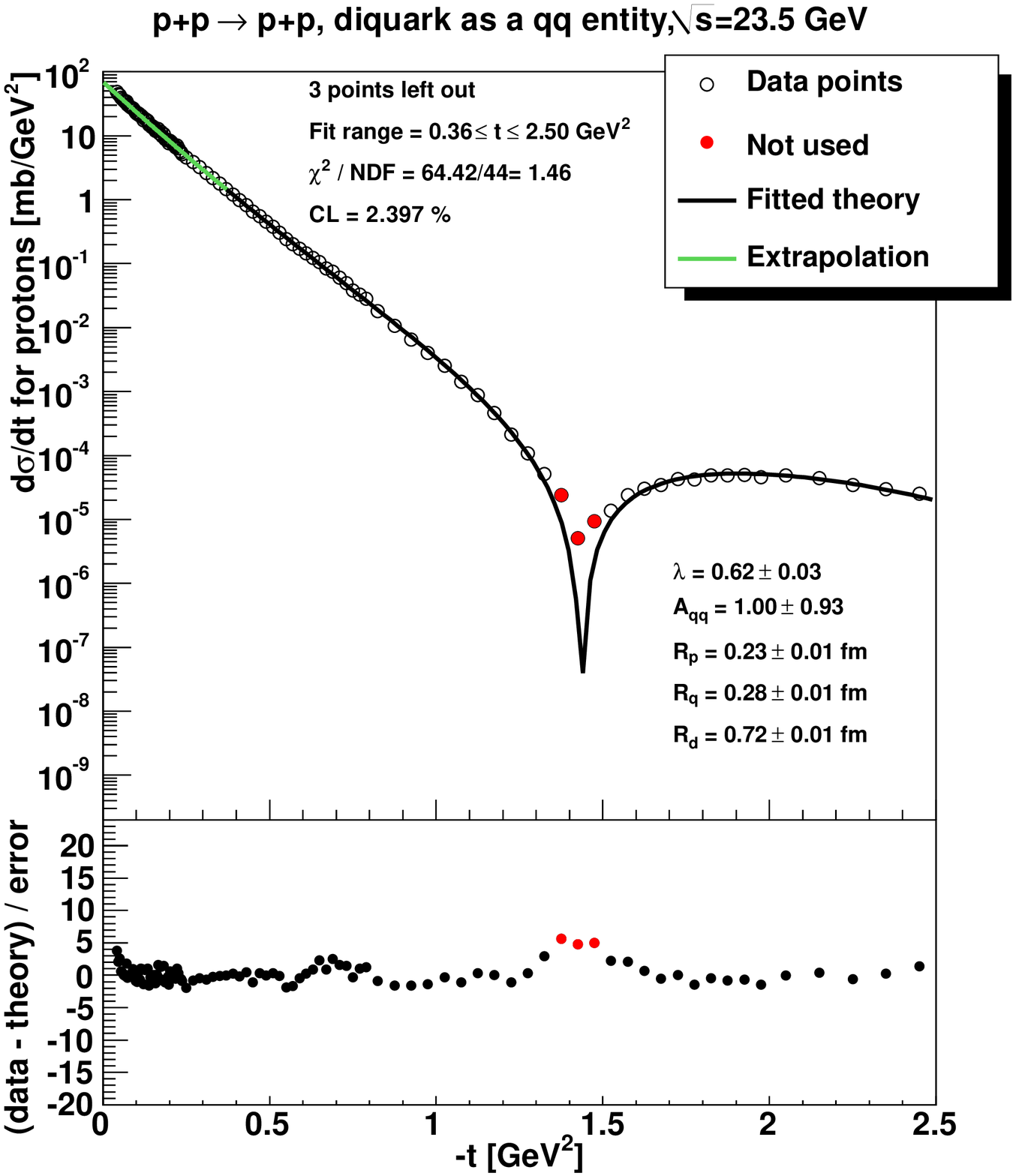}
        \centering
        \caption{
				The result of the fit at the 23.5 GeV ISR energy when the diquark is assumed to be a composite entity.  The confidence level is higher than 0.1\%, which means that the
                fit quality is acceptable.
                The fit was mode on the 0.36 - 2.5 GeV $\left|t\right|$ range according to ~\cite{Antchev:2011zz}.
                As the model is singular around the dip, 3 data points at the dip were left out from the fit.
                The parameter values are given with statistical errors. The systematic errors are not yet investigated and the effects of the correlation between
                model parameters are not yet determined, in this sense this fit result is still preliminary.
			}
		\label{qqfitfor23}
	\end{figure}

\begin{figure}[H]
    	\includegraphics[width=0.9\textwidth]{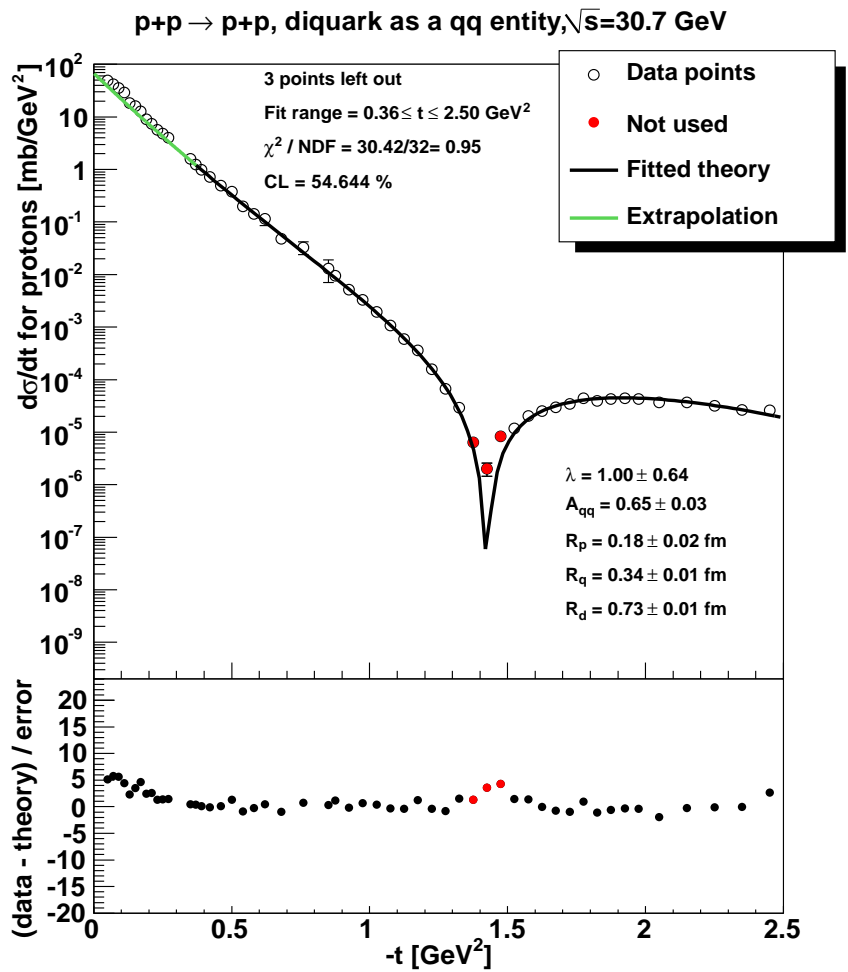}
        \centering
        \caption{Same as Fig \ref{qqfitfor23}, except that it is provided for $\sqrt{s}=$ 30.7 GeV.}
		\label{qqfitfor31}
	\end{figure}

\begin{figure}[H]
    	\includegraphics[width=0.9\textwidth]{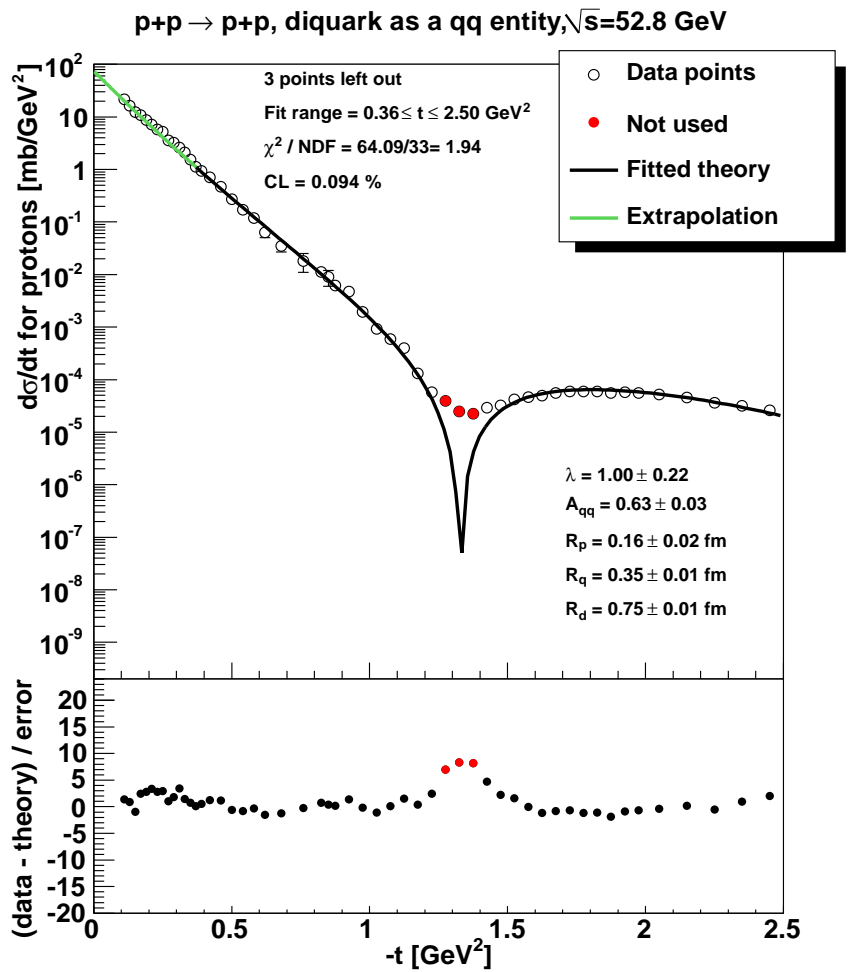}
        \centering
        \caption{Same as Fig \ref{qqfitfor23}, given for $\sqrt{s}=$ 52.8 GeV.}
		\label{qqfitfor53}
	\end{figure}

\begin{figure}[H]
		\includegraphics[width=0.9\textwidth]{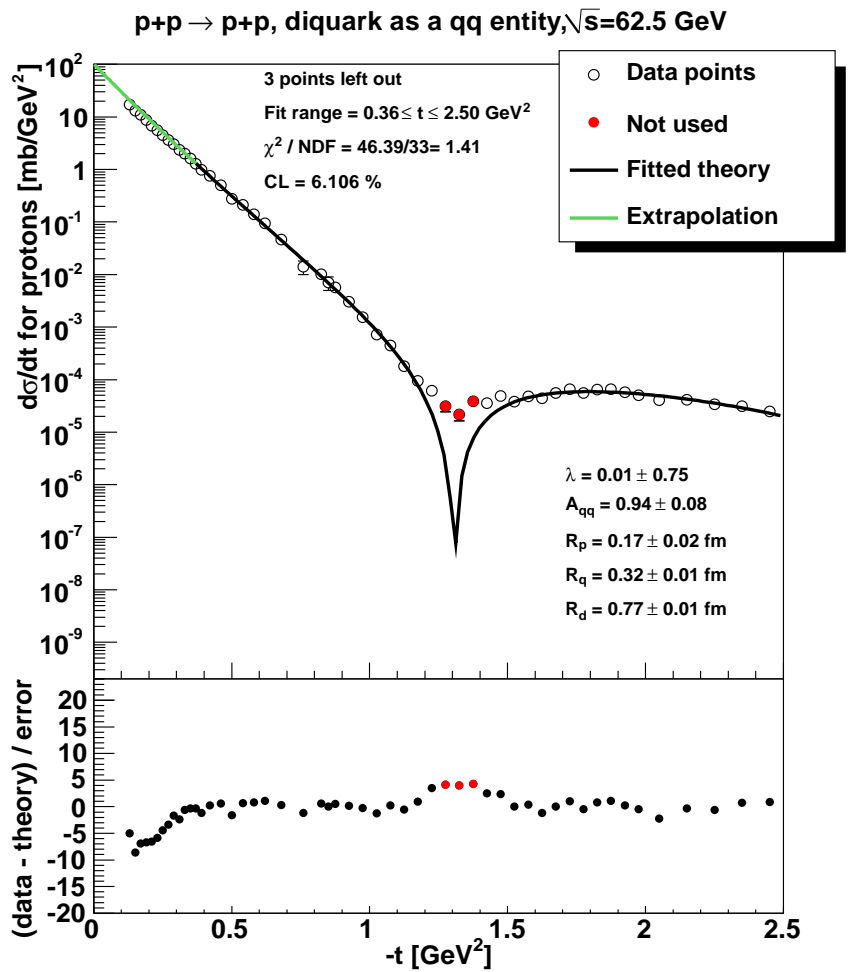}
		\centering
		\caption{Same as Fig \ref{qqfitfor23}, calculated for $\sqrt{s}=$ 62.5 GeV.}
		
		\label{qqfitfor62}
        \end{figure}

\begin{figure}[H]
		\includegraphics[width=0.9\textwidth]{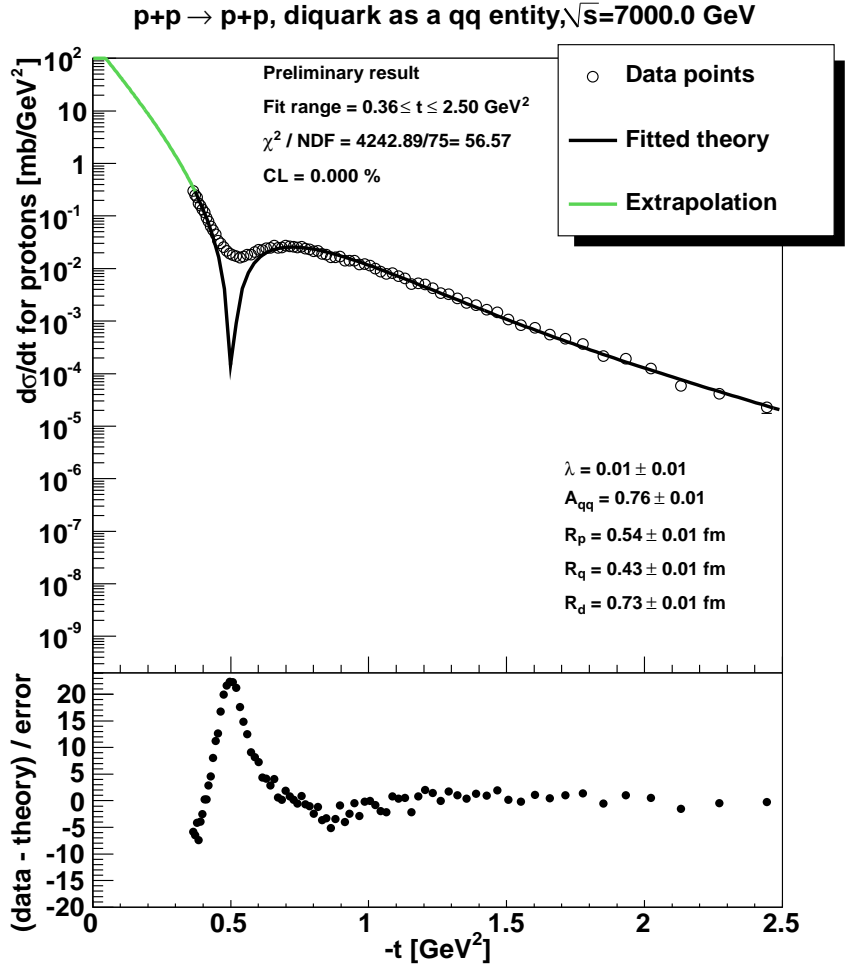}
		\centering
		\caption{
                Same as Fig \ref{qqfitfor23}, but for $\sqrt{s}=$ 7 TeV. Note that this fit is not acceptable since CL is below 0.1\%.
		}
		\label{qq_fittotem}
	\end{figure}

        In this section the MINUIT results are presented for the ISR \cite{Nagy:1978iw,Amaldi:1979kd}

and 
	TOTEM \cite{Antchev:2011zz} proton-proton elastic scattering
        data using the assumption that the diquark is a composite qq object. The results are illustrated on Fig. \ref{qqfitfor23}-\ref{qq_fittotem}. The confidence levels, and model parameters
        with their errors are summarized in Table \ref{tableqq}.
\vfill
\newpage

	\begin{table}
    	\begin{tabular}{|c|c|c|c|c|c|} \hline
			$\sqrt{s}$ [GeV]& 23.5 & 30.6 & 52.9 &  62.5 & 7000 \\ \hline\hline 
			$R_{p}$ [fm] & 0.23 $\pm$ 0.01 &  0.18 $\pm$ 0.02 &  0.16 $\pm$ 0.02 &  0.17 $\pm$ 0.02 &  0.54 $\pm$ 0.01 \\ \hline 
			$R_{q}$ [fm] & 0.28 $\pm$ 0.01 &  0.34 $\pm$ 0.01 &  0.35 $\pm$ 0.01 &  0.32 $\pm$ 0.01 &  0.43 $\pm$ 0.01 \\ \hline 
			$R_{d}$ [fm] & 0.72 $\pm$ 0.01 &  0.73 $\pm$ 0.01 &  0.75 $\pm$ 0.01 &  0.77 $\pm$ 0.01 &  0.73 $\pm$ 0.01 \\ \hline 
			$\lambda$ & 0.62 $\pm$ 0.03 &  1.00 $\pm$ 0.64 &  1.00 $\pm$ 0.22 &  0.01 $\pm$ 0.75 &  0.01 $\pm$ 0.01 \\ \hline 
			$A_{qq}$ & 1.00 $\pm$ 0.93 &  0.65 $\pm$ 0.03 &  0.63 $\pm$ 0.03 &  0.94 $\pm$ 0.08 &  0.76 $\pm$ 0.01 \\ \hline 
			$\chi^2/NDF$ & 64.42/44 & 30.42/32 & 64.1/33 & 46.4/33 & 4242.9/75 \\ \hline 
			$CL$ [\%] & 2.40 & 54.6 & 0.1 & 6.1 & 0.0  \\ \hline 
			$\sigma^{elastic}_{total}$ [mb] & 
			6.5  & 
			6.0  & 
			6.4  & 
			8.4  & 
			13.1 \\ \hline 
		\end{tabular}
		\caption{The overall fit quality and resulting parameters of the fit at the ISR energies including the LHC result at $7$ TeV. The 
				diquark is a assumed to be a qq entity.
				The obtained parameter values are given with statistical errors. The systematic errors are not yet investigated and the effects of the correlation between
                model parameters are not yet determined, so these results are preliminary.}
		\label{tableqq}
	\end{table}

%\newpage

\subsection{Total cross sections in the composite diquark model}
The total inelastic cross sections for the quark-quark, quark-diquark and diquark-diquark
subcollisions were analysed according to formula (\ref{inelastic cross sections}). The detailed results are
collected in Table \ref{table:tableinelasticsigma}, while the average ratios for the described ISR energies are

    \begin{table}
        \begin{tabular}{|c|c|c|c|c|c|} \hline
            $\sqrt{s}$ [GeV]& 23.5 & 30.6 & 52.9 &  62.5 & 7000 \\ \hline\hline
            $\sigma_{qd}/\sigma_{qq}$ & 1.92 $\pm$ 0.08 &  1.93 $\pm$ 0.01 & 1.93  $\pm$ 0.01 & 1.92 $\pm$ 0.01 &  1.87 $\pm$ 0.01 \\ \hline
            $\sigma_{dd}/\sigma_{qq}$ & 3.64 $\pm$ 0.34 & 3.66  $\pm$ 0.03 & 3.67  $\pm$ 0.03 & 3.61 $\pm$ 0.05 &  3.38 $\pm$ 0.02 \\ \hline
        \end{tabular}
        \caption{The ratios of the total inelastic cross sections for the quark-quark, quark-diquark and 
		diquark-diquark processes for the ISR and 
 		LHC energies using the composite diquark hypothesis.}
        \label{table:tableinelasticsigma}
    \end{table}

\begin{equation}
\sigma_{qq}:\sigma_{qd}:\sigma_{dd} = 1 :  (1.93 \pm 0.03) : (3.64 \pm 0.1), 
\label{ISRratios}
\end{equation}
which is close to the ideal 1: 2 : 4 ratio, confirming the assumption of having two quarks inside 
the diquark, amended with some shadowing which is 3.5\% and 9\% respectively. 
At 7 TeV the ratios are slightly different from (\ref{ISRratios})  
\begin{equation}
	1 : (1.87 \pm 0.01) : (3.38 \pm 0.02),
\end{equation}
which shows that the shadowing is stronger, 6.5\% and 16\% percent respectively.\par

The total elastic scattering cross sections were also determined, as given in Table \ref{tableqq}.
Note, that the errors on the total cross-section have not yet been obtained reliably at this point.
The reason is that the errors of total cross-sections are very strongly correlated with the errors
of $A_{qq}$. However the current fits cannot determine precisely the value of this parameter: its
nearly 100 \% relative error indicates the approximate insensitivity of the results on this value.
Hence one has to study the possibility of fixing this and also the parameters like  $\lambda$ 
to reasonable values and to see if in such scenarios the quality of the results remains the same or not.These studies and the final values of the total elastic scattering cross-sections will be 
submitted for a separate publication.

\newpage
\section{Conclusion and outlook}
\label{sec:conclusion}
	A systematic study of fit quality as well as the fit parameters under similar 
    circumstances has been performed for the Bialas - Bzdak model in a wide energy
 	range from ISR to LHC energies. The model gives a good description of the ISR data, which means that the CL is acceptable on the
	ISR energies if 3 data points at the dip were left out from the fit. The total proton-proton cross section (``size'' of the proton) 
	clearly seems to grow with energy but the model fails at this energy domain as  CL is not acceptable at $7$ TeV. This preliminary result does not include systematics, and
	the correlation between the fit parameters are still under 
	investigation.

    An important shortcoming of the quark-diquark model of protons is that it ignores the real part
    of the elastic scattering amplitude. This leads to a singular behaviour at the diffractive minimum,  which is apparently a more and more serious model limitation with increasing the energy of the p+p collisions.

    The evaluated ratios of the quark-quark, quark-diquark and diquark-diquark total inelastic cross-sections were found to deviate more and more from the ideal 1 : 2 : 4 ratio with increasing energies. In the ISR energy range the deviations from this ideal value were not yet significant, indicating lack of significant shadowing effects. However at the current LHC energy of $\sqrt{s} = 7$ TeV, a significant decrease compared
    to these ideal ratios were found, which possibly may indicate an increased role of shadowing
    at CERN LHC energies.

    The final conclusions of this study will be summarized separately and submitted for a publication.
    The current status corresponds to the level of our understanding around September 2011, at the
    time of the 2011 Summer School on Diffraction  in Heidelberg, where these results were first presented.

    Finally let us note that the TOTEM Collaboration extended recently the measurement of
    the differential elastic p+p scattering cross-sections to low values of $|t|$
    in ref.~\cite{Antchev:2011vs}, allowing one
    to extrapolate to the optical point at $t=0$ and to determine the total elastic and the total
    scattering cross-sections of p+p collisions at $\sqrt{s} = 7$ TeV for the first time, but
    these data were yet not utilized in our analysis.

    As a final remark, at the time of the completion of this conference contribution it is
    inspiring to see the great theoretical interest in the differential elastic cross-section measurement of TOTEM at LHC, as evidenced by the increasing amount of theoretical interpretations and
    successfull descriptions of certain aspects of these data. As our contribution is not intended
    to be a review on the interpretation of TOTEM data, we just would like to call attention to
    some of the most interesting approaches of describing these measurements as evidenced in
    refs. ~\cite{Fagundes:2011wn,Martin:2011gi,Wibig:2011iw,Uzhinsky:2011qu,Donnachie:2011aa,Fagundes:2011zx,Block:2012yx,Ryskin:2012ry}.

\vfill
\newpage

\section{Acknowledgements}
{ T. Cs\"org\H{o} would like to thank prof. R. J. Glauber for 
valuable discussions at the initial phase of this project,  
and for his kind hospitality at Harvard University.
F.N. would like to thank A. Ster and M. Csan\'ad for their valuable help 
with the CERN MINUIT multi-parameter optimization. 

This research was partially supported
by the Hungarian American Enterprize Scholarship Fund (HAESF) and by the Hungarian OTKA
grant NK 101438.
}

\newpage
\section{Appendix}
	Two of the Dirac $\delta$ functions in (\ref{elsoegyenlet}) induce the following transformation in the transverse 
	diquark and quark position variables
	\begin{equation}
		\vec{s_d}=-\lambda \vec{s_q},\, \vec{s_d}'=-\lambda \vec{s_q}'.
	\end{equation}
	Hence four Gaussian integration remain, which lead us to the following result
	\begin{align}
		\frac{4v^2}{\pi^2} & \int{\text{d}^2s_q \text{d}^2s_q' e^{-2v\left(s_q^2+s_q'^2\right)}
			e^{-c_{qq}\left(b-s_q+s_q'\right)^2} e^{-c_{qd}\left(b-s_q+s_d'\right)^2}\times} \\
			& \times e^{-c_{dq}\left(b-s_d+s_q'\right)^2} e^{-c_{dd}\left(b-s_d+s_d'\right)^2}=\frac{4v^2}{\Omega}e^{-b^2\frac{\Gamma}{\Omega}} \notag,
	\end{align}
	where the coefficients $c_{ab}$ are abbrevations, and
	\begin{align}
		\Omega&=\left[4v + \left(1+\lambda\right)^2\left(c_{qd}+c_{dq}\right) \right]\left[v+c_{qq} + \lambda^2 c_{dd}\right]+\\
		&+\left(1-\lambda\right)^2\left[v \left(c_{qd}+c_{dq}\right)+\left(1+\lambda\right)^2c_{qd}c_{dq}\right] \notag,
	\end{align}
	while
    \begin{align}
        \Gamma&=\left[4v + \left(1+\lambda\right)^2\left(c_{qd}+c_{dq}\right) \right]\left[v \left(c_{qq}+c_{dd}\right)+\left(1+\lambda\right)^2c_{qq}c_{dd}\right]+\\
        &+\left[4v + \left(1+\lambda\right)^2\left(c_{qq}+c_{dd}\right) \right]\left[v \left(c_{qd}+c_{dq}\right)+\left(1+\lambda\right)^2c_{qd}c_{dq}\right] \notag.
    \end{align}

\end{document}